\documentstyle[12pt]{article}
\begin{document}
\title{A New Spin on the Dirac Electron}
\author{B.G.Sidharth \\
Centr for Applicable Mathematics and Computer Sciences\\
B.M.Birla Science Centre,Hyderabad, 500463,India}
\date{}
\maketitle
\begin{abstract}
We re-examine the non Hermitian position coordinate of Dirac's equation, in
the light of his own insights and conclude that this, and the Dirac equation
itself is symptomatic of an underlying Noncommutative Geometry.
\end{abstract}
\section{Introduction}
Albert Einstein had once observed "you know, it would be
sufficient to really understand the electron" \cite{ra}. At the
turn of the Twentieth Century several valiant attempts were made
to model the electron in terms of Classical Electrodynamics, but
all these attempts ultimately failed\cite{r1}. As Wheeler pointed
out\cite{r2}, the real challenge was to introduce into the theory
the purely Quantum Mechanical spin half of the electron. The first
successful theory of the electron emerged with Dirac's equation,
which also at the same time brought about the unification of
Quantum Theory with the Special Theory of Relativity.\\ However
the Dirac theory also encountered several problems. One of these
was that the position coordinate turned out to be complex or Non
Hermitian, and as we will see, a related problem, namely that the
velocity of the electron turned out to be the velocity of light.
Dirac himself recognized the reason for all this. He
remarked\cite{r3}, "... since the theoretical velocity in the
above conclusion is the velocity at one instant of time while
observed velocities are always average velocities through
appreciable time intervals...", and again, "To measure the
velocity we must measure the position at two slightly different
times and then divide the change of position by the time interval.
(It will not do to measure the momentum and apply a formula, as
the ordinary connection between velocity and momentum is not
valid.) In order that our measured velocity may approximate to the
instantaneous velocity, the time interval between the two
measurements of position must be very short and hence these
measurements must be very accurate. The great accuracy with which
the position of the electron is known during the time-interval
must give rise, according to the principle of uncertainty, to an
almost complete indeterminacy in its momentum. This means that
almost all values of the momentum are equally probable, so that
the momentum is almost certain to be infinite. An infinite value
for a component of momentum corresponds to the value $\pm c$ for
the corresponding component of velocity."\\ This realization,
which highlights the limitation of space time points in Quantum
Theory highlights the fact that we have to deal instead, with
minimum space time intervals, within which there are negative
energy solutions and the zitterbewegung type of unphysical
effects. On the other hand negative energy components of the Dirac
bi-spinor are negligible outside the Compton scale. Thus the
averaging prescribed by Dirac eliminates these components and
gives us back a physical theory in terms of Hermitian operators
(Cf.\cite{r4}). This realization is the seed of what in recent
years has been termed a Non Commutative Geometry.\\
\section{Positive and Negative Energy Solutions}
Let us consider in a little more detail\cite{r4} the implications of Dirac's
averaging over the Compton scale.\\
We consider for simplicity, the free particle Dirac equation. The
solutions are of the type,
\begin{equation}
\psi = \psi_A + \psi_S\label{e25}
\end{equation}
where
$$
\psi_A =   e^{\frac{\imath}{\hbar} Et} \ \left(\begin{array}{l}
                                          0 \\ 0 \\ 1 \\ 0
                             \end{array}\right) \mbox{ or } \ e^{\frac{\imath}{\hbar} Et}
                          \ \   \left(\begin{array}{l}
                                 0 \\ 0 \\ 0 \\ 1
                              \end{array}\right) \mbox{ and }
$$
\begin{equation}
\label{e26}
\end{equation}
$$
\psi_S =  e^{-\frac{\imath}{\hbar} Et} \ \left(\begin{array}{l}
                                 1 \\ 0 \\ 0 \\ 0
                               \end{array}\right) \mbox{ or } e^{-\frac{\imath}{\hbar} Et}
                            \ \   \left(\begin{array}{l}
                                 0 \\ 1 \\ 0 \\ 0
                               \end{array}\right)
$$
denote respectively the negative energy and positive energy solutions. From
(\ref{e25}) the probability of finding the particle in a small volume
about a given point is given by
\begin{equation}
| \psi_A + \psi_S|^2 = |\psi_A|^2 + |\psi_S|^2 +
(\psi_A \psi_S^* + \psi_S \psi_A^*)\label{e27}
\end{equation}
Equations (\ref{e26}) and (\ref{e27}) show that the negative energy and
positive energy solutions form a coherent Hilbert space and so the
possibility of transition to negative energy states exists. This difficulty
however can be overcome by the well known Hole theory which uses the Pauli exclusion
principle, and is described in many standard books on Quantum Mechanics.\\
However the last or interference term on the right side of (\ref{e27}) is like the
zitterbewegung term. When we remember that we really have to consider
averages over spacetime intervals of the order of $\hbar/mc$ and
$\hbar/mc^2$, this term disappears and effectively the negative energy
solutions and positive energy solutions stand decoupled in what is now
the physical universe. In other words, the Hole theory and other artifices
of point space time theory are circumvented if, self consistently we use
space time intervals instead of points.\\
The spirit of Dirac's average spacetime intervals rather than spacetime points
has received attention over the years in the form of minimum
spacetime intervals-- from the work of Snyder and Schild to Quantum Superstring
theory \cite{r5,r6,r7,r8,r9,r10,r11}. In modern
language, it is symptomatic of a Non commutative spacetime geometry which
again brings out the nature of the mysterious Quantum Mechanial spin  \cite{r12,r13}.
This is what we will briefly examine.
\section{The Non commutative Structure}
Very early on, Newton and Wigner \cite{r14} showed that the correct physical coordinate operator
is given by
\begin{equation}
x^k = (1 +\gamma^0 ) \frac{p_0^{3/2}}{(p_0+\mu)^{1/2}} \left(-\frac{\imath \partial}
{\partial p_k}\right) \frac{p_0^{-1/2}}{(p_0+\mu)^{1/2}} P\label{e1}
\end{equation}
where $P$ is a projection operator eliminating negative energy components
and the gammas denote the usual Dirac matrices.\\
To appreciate the significance of (\ref{e1}), let us consider the case of
spin zero. Then (\ref{e1}) becomes
\begin{equation}
x^k = \imath \frac{\partial}{\partial p_k} + \frac{1}{8\pi} \int
\frac{\exp (-\mu |(x-y|)}{|x-y|} \frac{\partial}{\partial y} dy\label{e3}
\end{equation}
The first term on the right side of (\ref{e3}) denotes the usual position operator,
but the second term represents an imaginary part, which has an extension $\sim 1/\mu$,
the Compton wavelength, exactly as in the case of the Dirac electron.\\
Returning to Dirac's treatment \cite{r3}, the position coordinate is given by
\begin{equation}
\vec x = \frac{c^2pt}{H} + \frac{1}{2} \imath c \hbar (\vec \alpha - c\vec p
H^{-1}) H^{-1} \equiv \frac{c^2p}{H}t + \hat x\label{e4}
\end{equation}
$H$ being the Hamiltonian operator and $\alpha$'s the non-commuting Dirac
matrices, given by
$$\vec \alpha = \left[\begin{array}{l}
\vec \sigma \quad 0\\
0 \quad \vec \sigma
\end{array}
\right]
$$
The first term on the right hand side of (\ref{e4}) is the usual (Hermitian) position. The
second term of $\vec x$ is the small oscillatory term of the order of the Compton
wavelength, arising out of zitterbewegung effects which averages out to zero.
On the other hand, if we were to work with the (non Hermitian) position operator in
(\ref{e4}), then we can easily verify that the following Non-commutative geometry
holds,
\begin{equation}
[x_\imath , x_j] = \alpha_{\imath j} l^2\label{e5}
\end{equation}
where $\alpha_{\imath j} \sim 0(1)$.\\
The relation (\ref{e5}) shows on comparison with the position-momentum commutator
that the coordinate $\vec x$ also behaves like a "momentum". This can be
seen directly from the Dirac theory itself where we have
\begin{equation}
c\vec \alpha = \frac{c^2 \vec p}{H} - \frac{2\imath}{\hbar} \hat x H\label{e6}
\end{equation}
In (\ref{e6}), the first term gives the usual momentum. The second term is the extra
"momentum" $\vec {\hat p}$ due to the relations (\ref{e5}).\\
Infact we can easily verify from (\ref{e6}) that
\begin{equation}
\vec {\hat p} = \frac{H^2}{\hbar c^2} \hat x\label{e7}
\end{equation}
where $\hat x$ has been defined in (\ref{e4}).\\
Let us now see what the angular momentum $\sim \vec x \times \vec p$ gives
at the Compton scale. Using (\ref{e4}), we can
easily show that
\begin{equation}
(\vec x \times \vec p )_z = \frac{c}{E} (\vec \alpha \times \vec p )_z = \frac{c}{E}
(p_2 \alpha_1 - p_1\alpha_2)\label{e8}
\end{equation}
where $E$ is the eigen value of the Hamiltonian operator $H$. The right side of
(\ref{e8}) is a super position of the $\alpha$'s which again contain the Pauli
$\sigma$ matrices. This shows that at the Compton scale, the angular momentum
leads to the "mysterious" Quantum Mechanical spin.\\
It may be mentioned that Zakruzewski \cite{r46} deduced from a different point
of view that spin implies non commutativity. On the other hand, the zitterbewegung
contribution to spin has been shielded by Barut and coworkers and
Hestenes (Cf.ref.\cite{r4} and several references therein).\\
In the above considerations, we started with the Dirac equation and deduced
the underlying Noncommutative geometry of spacetime. Interestingly, starting
with Snyder's Non commutative geometry, based solely on Lorentz invariance and a
minimum spacetime length, at the Compton scale,
$$[x,y] = \frac{\imath l^2}{\hbar} L_z etc.$$
that is, in effect starting with (\ref{e5}), it is possible to deduce the relations
(\ref{e8}),(\ref{e7}) and the Dirac equation itself as has been shown in
detail elsewhere \cite{r10,r46,r16,r17}.\\
We have thus established the correspondence between considerations starting from
the Dirac theory of the electron and Snyder's (and subsequent) approaches based
on a minimum spacetime interval and Lorentz covariance.
We will now show using Nelson's analysis that that the above non commutativity is also symptomatic of an underlying
stochastic behaviour.
\section{The Stochastic Underpinning}
In Nelson's approach\cite{r18,r19},
there is a double Weiner process arising from the fact that the forward and
backward time derivatives,
\begin{equation}
\frac{d}{dt^+}, \quad \frac{d}{dt^-}\label{e11}
\end{equation}
are unequal. Let us consider first the problem in one dimension (Cf.\cite{r19}))
we have
\begin{equation}
\frac{d_+}{dt} x(t) = b_+ \quad , \quad \frac{d_-}{dt}x(t) = b_-,\label{e12}
\end{equation}
From (\ref{e12}) we define two new velocities
\begin{equation}
V = \frac{b_+ + b_-}{2} \quad ; \quad U = \frac{b_+ - b_-}{2}\label{e13}
\end{equation}
It may be pointed out that in general $U$, given in (\ref{e13}) vanishes while
$V$ gives the usual velocity. It is now possible to introduce a complex velocity
\begin{equation}
\vee = V - \imath U\label{e14}
\end{equation}
From (\ref{e14}) it appears that the coordinate $x$ goes over to a complex
coordinate
\begin{equation}
x \to x + \imath x'\label{e15}
\end{equation}
This is also true for the Dirac equation (\ref{e4}). Infact it can be shown that this leads to the special relativistic metric in
$1+1$ dimensions \cite{r4}.\\
Following this line of reasoning, in the usual theory, as is well known we work with (\ref{e14}) to deduce the
Schrodinger equation. This could be done in three
dimensions also.\\
But let us now look upon (\ref{e15}) from a different angle and ask,
"Can we generalise
(\ref{e15}) itself to the three dimensional case?" It is known \cite{r20} that
in this case, surprisingly,
\begin{equation}
(1, \imath ) \to (I, \vec \sigma )\label{e16}
\end{equation}
where $I$ is the unit $2 \times 2$ matrix and $\vec \sigma$ are
the Pauli matrices. We get the Lorentz invariant metric at the
same time.\\ Equation (\ref{e16}) gives a quarternionic
description (Cf.\cite{r20}). It would lead to a two component
neutrino type equation. However this $2 \times 2$ description does
not preserve spatial and time reflections, which are necessary in
physical theories. If we incorporate these reflections also then
it is known that \cite{r21} we get the $4 \times 4$ Dirac
description and the non commutative geometry \cite{r12}
\begin{equation}
[x_\imath , x_j] = l^2 \Theta_{\imath j}\label{e17}
\end{equation}
where $l$ represents a length scale. Equation (\ref{e17}) is,
ofcourse, the same as (\ref{e5})! All this need not be surprising
- equation (\ref{e4}) represents the zitterbewegung effects due to
the interference of the negative and positive energy solutions.
The positive and negative time derivatives (\ref{e11}) of the
double Weiner process described above represent exactly the
positive and negative energy interference effects, contained in
the term $U$ of (\ref{e13}). (It is these interference terms
which, even in the non relativistic Nelsonian theory lead to the
Quantum Mechanical Schrodinger equation).\\ The important point to
note is that whenever we have a complex space coordinate as in
(\ref{e15}), then the generalization to three dimensions, infact
leads to (\ref{e16}) and non commutativity.\\ There is another
very interesting and apparently different situation where complex
coordinates are used - this is in the derivation of the
Kerr-Newman metric \cite{r22,r23,r24}. In this case we consider
the Maxwell equations,
\begin{equation}
\vec \nabla \times \vec W = \imath {\vec W}, \vec \nabla \cdot \vec W = 0\label{e18}
\end{equation}
where
$$\vec W = \vec E + \imath \vec B$$
$\vec E$ and $\vec B$ being the usual electric and magnetic components of the
electromagnetic field. The interesting point is that if we effect an imaginary
shift of the coordinates, then we obtain the Quantum Mechanical electron anomalous gyro magnetic ratio
$g = 2$ of the Kerr-Newman metric (Cf.\cite{r22} for details). Newman himself
found this association of the imaginary spatial shift with Quantum Mechanical
spin inexplicable \cite{r25}. However in the light of the above comments
the connection between the two becomes clear (Cf. also \cite{r26}).
Infact this leads to a model of the electron as a Kerr-Newman black hole, with
the naked singularity shielded by a fuzzyness induced by the non commutativity
(\ref{e17}) as was independantly confirmed by Nottale \cite{r27}.\\
For completeness we mention that interestingly, as is well known, the hydrodynamical formulation of Quantum Mechanics also leads
to equations similar to the Nelsonian formulation (Cf.\cite{r4} for a brief
review). Here also if we consider a one dimensional
laminar flow we get a velocity that is both solenoidal and
irrotational and satisfies
\begin{equation}
\vec \nabla \cdot \vec V = 0, \quad \vec \nabla \times \vec V = 0\label{e19}
\end{equation}
From (\ref{e19}) we can define a complex velocity potential by standard methods
which again leads to complex coordinates (\ref{e15}) and ultimately we end up
with (\ref{e17}).\\
We finally observe that a phase space approach based on relations like (\ref{e15}) has been worked
out in detail by Kaiser \cite{r28,r29}, though he does not follow the
route through (\ref{e16}), and therefore does not arrive at the above conclusions.
One can see that equation (\ref{e7}) also suggests this approach.
Nevertheless it is interesting to note that in this reverse approach in which
we introduce complex coordinates the zitterbewegung as manifested in the complex
coordinate $\hat x$ of (\ref{e4}) disappears.
\section{Conclusion}
Thus the purely Quantum Mechanical "mysterious" spin half is symptomatic of
non commutativity and vice versa. Interestingly, this is also symptomatic of
an underlying double Weiner or stochastic process.

\end{document}